\documentclass[conference]{IEEEtran}
\IEEEoverridecommandlockouts
\usepackage{amsmath,amssymb,amsfonts}
\usepackage{tabularx}
\usepackage{textcase}
\usepackage{booktabs, tabularx}
\usepackage{multirow}
\usepackage{array}
\usepackage[caption=false,font=normalsize,labelfont=sf,textfont=sf]{subfig}
\usepackage{textcomp}
\usepackage{stfloats}
\usepackage{url}
\usepackage{verbatim}
\usepackage{graphicx}
\usepackage{cite}
\usepackage{pifont}
\usepackage{lettrine}
\usepackage{mathtools}
\usepackage{hyperref}
\usepackage[export]{adjustbox}
\usepackage{subfig}    % for \subfloat macro
\usepackage[ruled, lined, linesnumbered, commentsnumbered, longend]{algorithm2e}
\newcolumntype{M}[1]{>{\centering\arraybackslash}m{#1}}
\newcolumntype{N}{@{}m{0pt}@{}}
\usepackage{makecell}

\usepackage[top=54pt, left=48pt, right=48pt, bottom=69pt]{geometry}

\begin{document}
	
	\title{Outlier Rejection for 5G-Based Indoor Positioning in Ray-Tracing-Enabled Industrial Scenario \\
		%{\footnotesize \textsuperscript{*}Note: Sub-titles are not captured in Xplore and
		%should not be used}
		%\thanks{This Project has received funding from the European Union’s Horizon 2020 research and innovation programme under the Marie Sklodowska-Curie grant agreement No 956670.}
	}
	
	\makeatletter
    \newcommand{\linebreakand}{%
    \end{@IEEEauthorhalign}
    \hfill\mbox{}\par
    \mbox{}\hfill\begin{@IEEEauthorhalign}
}
\makeatother
	
	\author{
    \IEEEauthorblockN{
        Karthik Muthineni\IEEEauthorrefmark{1}\IEEEauthorrefmark{2}, Alexander Artemenko\IEEEauthorrefmark{1}, Josep Vidal\IEEEauthorrefmark{2}, Montse Nájar\IEEEauthorrefmark{2}
    }
    \IEEEauthorblockA{\IEEEauthorrefmark{1} Corporate Sector Research and Advance Engineering, Robert Bosch GmbH, Renningen, Germany}
    \IEEEauthorblockA{\IEEEauthorrefmark{2} Department of Signal Theory and Communications, Universitat Politècnica de Catalunya (UPC), Barcelona, Spain}
    \IEEEauthorblockA{Email: \IEEEauthorrefmark{1}$\{${karthik.muthineni, alexander.artemenko}$\}$@de.bosch.com, \IEEEauthorrefmark{2}$\{${josep.vidal, montse.najar$\}$@upc.edu}}
}
	
	%\author{\IEEEauthorblockN{Karthik Muthineni\IEEEauthorrefmark{1},
    %Alexander Artemenko\IEEEauthorrefmark{2}, Josep Vidal Manzano\IEEEauthorrefmark{3} and
    %Montse Nájar\IEEEauthorrefmark{4}}
    %\IEEEauthorblockA{Corporate Sector Research and Advance Engineering,
    %Whichever University\\
    %Wherever\\
    %Email: \IEEEauthorrefmark{1}karthik.muthineni@de.bosch.com,
    %\IEEEauthorrefmark{2}alexander.artemenko@de.bosch.com,
    %\IEEEauthorrefmark{3}josep.vidal@upc.edu,
    %\IEEEauthorrefmark{4}montse.najar@upc.edu}}
	
	%\author{\IEEEauthorblockN{Karthik Muthineni}
	%	\IEEEauthorblockA{\textit{Corporate Sector Research and Advance Engineering} \\
	%		\textit{Robert Bosch GmbH}\\
	%		Renningen, Germany \\
	%		karthik.muthineni@de.bosch.com}
	%	\and
	%	\IEEEauthorblockN{Alexander Artemenko}
	%	\IEEEauthorblockA{\textit{Corporate Sector Research and Advance Engineering} \\
	%		\textit{Robert Bosch GmbH}\\
	%		Renningen, Germany \\
	%		alexander.artemenko@de.bosch.com}
	%	\linebreakand
	%	\IEEEauthorblockN{Josep Vidal Manzano}
	%	\IEEEauthorblockA{\textit{Department of Signal Theory and Communications} \\
	%		\textit{Universitat Politècnica de Catalunya (UPC)}\\
	%		Barcelona, Spain \\
	%		josep.vidal@upc.edu}
	%	\and
	%	\IEEEauthorblockN{Montse Nájar}
	%	\IEEEauthorblockA{\textit{Department of Signal Theory and Communications} \\
	%		\textit{Universitat Politècnica de Catalunya (UPC)}\\
	%		Barcelona, Spain \\
	%		montse.najar@upc.edu}
		
	%}
	
	\maketitle
	
	\begin{abstract}
		
		The precise and accurate indoor positioning using cellular communication technology remains to be a prerequisite for several industrial applications, including the emergence of a new topic of Integrated Sensing and Communication (ISAC). However, the frequently occurring Non-Line-of-Sight (NLoS) conditions in a heavy multipath dominant industrial scenario challenge the wireless signal propagation, leading to abnormal estimation errors (outliers) in the signal measurements taken at the receiver. In this paper, we investigate the iterative positioning scheme that is robust to the outliers in the Time of Arrival (ToA) measurements. The Iteratively Reweighted Least Squares (IRLS) positioning scheme formulated on the Least Squares (LS) is implemented to reject the outlier measurements and reweight the available ToA samples based on their confidence. Our positioning scheme is validated under 5G frequency bands, including the C-band (3.7~GHz) and the mmWave-band (26.8~GHz) in a Ray-Tracing enabled industrial scenario with different emulation setups.
		
	\end{abstract}
	
	
	\section{Introduction}
	The fifth generation (5G) cellular communication technology is designed to support new services from the consumer market and the industry verticals. To this end, using the 5G communication network for precise and accurate positioning has been studied extensively in recent years~\cite{FAN2022, WEN2021, OSSI2021, MUTHINENI2023}. Positioning using cellular communication technology has gained attention from industries because of the emergence of the new topic of Integrated Sensing and Communication (ISAC), which opens a wide variety of business opportunities for industry verticals, and positioning remains a prerequisite for ISAC~\cite{VISH2020}.
%The primary usage of the 5G network is defined around three main application domains, namely massive Machine Type Communications (mMTC), Ultra-Reliable and Low Latency Communications (URLLC), and enhanced Mobile Broadband (eMMB). However, the secondary usage of the 5G network for precise and accurate positioning has also been studied extensively in recent years.

The radio channel conditions existing inside industrial buildings will have a higher influence on the achievable indoor positioning accuracy. The presence of heavy metallic objects, concrete walls, and a dynamic environment in the form of moving forklifts creates the Non-Line-of-Sight (NLoS) condition between the base station and the User Equipment (UE), which gives rise to Multi-Path Components (MPCs). The Time-of-Arrival (ToA), Time Difference of Arrival (TDoA), and Angle-of-Arrival (AoA) estimation under a heavy MPCs dominant scenario lead to errors \cite{MA2019}, which are referred to as outliers in this paper. Position computation with outliers in the measurements can compromise the precision and accuracy requirements of the industry verticals. In such situations, new innovative techniques for identifying and eliminating measurements with outliers before computing the UE position become essential. 

We focus on using TDoA measurements for positioning because of its efficiency and ease of implementation. Several methods have been proposed by the research community to identify and reject the outliers from the raw ToA/TDoA measurements, e.g., \cite{Guvenc2009, WANG2017, LEE2021, AMAR2010}. However, \cite{Guvenc2009} and \cite{WANG2017} have a drawback because a priori information about the status of the path (Line-of-Sight (LoS)/NLoS) and the signal transmission time is needed. A recursive method of weighted least squares is implemented in \cite{LEE2021}. It requires a unique reference base station, with respect to which the TDoA measurements are computed. However, this approach can lead to inaccurate position estimates if the measurement at the reference base station is found to be an outlier. A reference-free TDoA-based positioning has been presented in \cite{AMAR2010}. However, it does not deal with the outlier rejection. 
%\IEEEpubidadjcol

To overcome the above limitations, in this paper, similar to \cite{MARCUS2022}, we implement the Iteratively Reweighted Least Squares (IRLS) positioning technique, which is robust to the outliers. Instead of using a fixed base station as a reference in TDoA measurements, the IRLS positioning uses every base station as a reference once to obtain distinct possible position estimations. After that, a weighting criterion is employed to give less weight or reject the outlier measurements and assign more weight to those with higher confidence, resulting in a weighted average position estimate with higher accuracy. 

The contribution of this paper is to provide directions to the industry verticals about the achievable indoor positioning accuracy in industrial scenarios using 5G mobile communication technology operating at C-band and mmWave-band. To achieve this, we choose the production hall of the Bosch industry in Germany as the industrial scenario under investigation. The physical environment of the production hall is scanned with the 3D laser scanners and a Ray-Tracing enabled 3D model of the production hall is created. The efficiency of the IRLS positioning algorithm is evaluated under the C-band (3.7 GHz) and the mmWave-band (26.8 GHz) in the Ray-Tracing model of the real industry. Moreover, we elucidate the best choice of the 5G band for indoor positioning in industrial scenarios.

The rest of the paper is organized as follows: Section \ref{Problem Statement} presents the problem statement. In Section \ref{IRLS}, the steps and procedures for the IRLS positioning algorithm are explained in detail. Section \ref{Channel emulation} describes the industrial scenario, propagation environment, and channel emulation settings. The performance of the IRLS algorithm and the achievable indoor positioning accuracy are derived in Section \ref{Discussion}. Finally, Section \ref{Conclusion} provides the concluding remarks.

\section{Problem Statement}
\label{Problem Statement}
We consider a 2-dimensional (2D) layout, in which a UE with an isotropic antenna is placed at the point indicated with $\{\textrm{q}_{ue} = (x_{ue} \,, y_{ue})\,,\,\textrm{q}_{ue} \in \mathbb{R}^2\}$. The positioning infrastructure consists of $\mathrm{N}$ base stations each with an isotropic antenna placed at the points indicated with $\{\textrm{q}_{bs} =~(x_{bs} \,, y_{bs})\,, \,\textrm{q}_{bs} \in \mathbb{R}^2\,, \,bs \in \{1\,,\,2\,,\,.\,.\,.\,,\,\mathrm{N}\} \}$. We assume that the clocks of the base stations are synchronized and the signal transmissions from the base stations follow a sequential order with a period of $\delta$ = 10 ms between two base station transmissions. The signal received by the UE from one base station in an NLoS-dominant multipath environment is represented as
\begin{equation}
r_{bs}(t) = \sum_{j=1}^{M_{bs}} A_{bs}^j \cdot s(t \,- \tau_{bs}^j) + n_{rx}, \,\,\,\,\,bs \in \{1\,,\,2\,,\,.\,.\,.\,,\,\mathrm{N}\}
\label{eq1}
\end{equation}
where $s(t)$ is the transmitted signal, $M_{bs}$ represents the total MPCs received from the base station $bs$, and the amplitude as well as the ToA of the $j$-th MPC are given by $A_{bs}^j$ and $\tau_{bs}^j$. The last term $n_{rx}$ represents the noise at the receiver. To replicate a realistic scenario, a bias caused by the paths is added to the emulated MPCs. The ToA of the MPC is estimated from the maximum absolute value of the convolution between the channel impulse response and the raised cosine pulse given by
\begin{equation}
h(t) = \frac{1}{T} sinc\left(\frac{t}{T}\right)\frac{cos\left(\frac{\pi\beta t}{T}\right)}{1-\left(\frac{2\beta t}{T}\right)^2},
\label{eq2}
\end{equation}
where $T$ is the symbol period and $\beta$ is the roll factor dependent on the bandwidth of the C-band or the mmWave-band. In addition, $n_{rx}$ is the noise affecting the emulated MPC. Therefore, the estimation of ToA is affected by the Gaussian random noise of the variance $\sigma^2$ \cite{HUANG2015}
%the noise added to the emulated MPCs corresponds to a Gaussian noise with zero mean and variance based on the Cramer Rao Lower Bound (CRLB) of ToA estimation that depends on the signal bandwidth \cite{HUANG2015}
\begin{equation}
\sigma^2 = \frac{c^2}{(2 \pi B)^2 \cdot t_{s} \cdot B \cdot SNR},
\label{eq3}
\end{equation}
where $B$ represents the signal bandwidth, $t_s$ is the signal time period, and $SNR$ is the Signal-to-Noise Ratio (SNR) recorded at the UE. In this paper, the first-arriving MPC to the UE from each of the base stations is used for computing the position. The ToA associated with the first-arriving MPC can be defined as
\begin{equation}
\tau_{bs}^1 = \frac{1}{c}\left\{\sqrt{(x_{ue} - x_{bs})^2 + (y_{ue} - y_{bs})^2} \right\}.
\label{eq4}
\end{equation}
where $c$ defines the speed of the light. Therefore, the problem to solve is, given the inaccurate ToAs of the first-arriving MPCs, how to estimate the UE position? 

\begin{algorithm}[t]
	\SetKwFunction{isOddNumber}{isOddNumber}
	% \SetKwInput{Input}{Input}
	% \SetKwInput{Output}{Output}
	\SetKwInOut{KwIn}{Input}
	\SetKwInOut{KwOut}{Output}
	
	\KwIn{ToAs of first-arriving MPCs $\tau_{n}^1$, transmission period~$\delta_{ne}$, base station coordinates $(x_{bs}\,, y_{bs})$, maximum acceptable uncertainty $u_{max}$, threshold $\epsilon$}
	\KwOut{Weights of the base stations $w_{norm,n}$, weighted average position estimate $\hat{\textrm{Q}}_{\textrm{WA}}$}
	
	\For{$i \leftarrow 0$}{compute position estimates $\hat{\textrm{q}}$ using (\ref{eq5}) - (\ref{eq7}) \\ Initial estimate of $\hat{\textrm{Q}}_{\textrm{WA}}$ using (\ref{eq8})}
	
	\While{$S$ $>$ $\epsilon$}
	{$i \leftarrow i + 1$ \\
	Calculate uncertainty $u_e$ using (\ref{eq10})\\
	Update weights of the base stations using (\ref{eq11})\\
    Get normalized weights $\Bar{w}_{norm, n}$ using (\ref{eq9})\\
    Get new weighted average position estimate $\hat{\textrm{Q}}_{\textrm{WA}}$ using (\ref{eq8})\\
    Compute convergence $S$ using (\ref{eq12})}

	$\hat{\textrm{Q}}_{\textrm{IRLS}} \leftarrow \hat{\textrm{Q}}^i_{\textrm{WA}}$
	
	\textbf{return} $\hat{\textrm{Q}}_{\textrm{IRLS}}\,, w_{norm}$

	%\tcc{For odd elements in the list, we add 1, and for even elements, we add 2.
	%	After the loop, all elements are even.}
	%\For{$i \leftarrow 0$ \KwTo $n-1$}{
	%	\eIf{$\isOddNumber(a_i)$}{
			
	%		$newList.append(a_i + 1)$ \tcp*[f]{Some thought-provoking comment.}
	%	}{
	%		\tcp{Another comment}
	%		$newList.append(a_i + 2)$
	%	}
	%}
	
	\caption{IRLS Positioning Algorithm}
        \label{alg1}
\end{algorithm}

\section{IRLS Positioning Algorithm}
\label{IRLS}
Taking random base station as a reference $e$, the $\mathrm{N}-1$ TDoAs can be computed as
\begin{equation}
TDoA_{ne} = |(\tau_{n}^1 - \tau_{e}^1) - \mathrm{\delta}_{ne}|, \,\,\,\,\,n \in \{1\,,\,2\,,\,.\,.\,.\,,\,\mathrm{N}\} \setminus \{e\}
\label{eq5}
\end{equation}
where $\tau^1$ represents the ToA of the first arriving MPC and $\delta_{ne}$ is the period between the two base station signal transmissions. To obtain position, (\ref{eq5}) can be multiplied with the speed of the light ${c}$, which yields difference in distance ($\Delta{d}$) as
 \begin{equation}
\begin{split}
(\Delta{d})_{ne} = \Big[\sqrt{(x_{ue}-x_n)^2 + (y_{ue}-y_n)^2}\\-\sqrt{(x_{ue}-x_e)^2 + (y_{ue}-y_e)^2}\Big],
\label{eq6}
\end{split}
\end{equation}
Solving $\mathrm{N}-1$ equations for base stations of the format shown in (\ref{eq6}) with the Least Squares (LS) technique, we obtain the position of the UE. The idea of LS is to find a point $(x_{ue}, y_{ue})$, so that it minimizes the squared error function as%$\hat{\textrm{\textbf{p}}}$ $(x_{ue}\,, y_{ue})$ 
\begin{equation}
\begin{split}
LS = \sum_{n \in bs\setminus \{e\}}^\textrm{N}\Big((\Delta{d})_{ne} - \Big[\sqrt{(x_{ue}-x_n)^2 + (y_{ue}-y_n)^2}\\-\sqrt{(x_{ue}-x_e)^2 + (y_{ue}-y_e)^2}\Big]\Big)^2,
\label{eq7}
\end{split}
\end{equation}
In (\ref{eq7}), if the ToA measurement recorded at the reference base station $(x_{e}, y_{e})$ is found to be an outlier, it will degrade the overall positioning accuracy. To overcome this drawback, we use every base station as a reference once to obtain $\mathrm{N}$ different position estimates $\hat{\mathrm{Q}} = [\hat{\mathrm{q}}_1\,, \hat{\mathrm{q}}_2\,,\,.\,.\,.\,,\, \hat{\mathrm{q}}_\textrm{N}]$. We then assign weights to each of the position estimates and compute the weighted average position estimate given by
\begin{equation}
	\hat{\textrm{Q}}_{\textrm{WA}} = \sum_{n = 1}^{\textrm{N}} \Bar{w}_{norm, n}\cdot \hat{\textrm{q}}_{n},
	\label{eq8}
\end{equation}
where $\Bar{w}_{norm, n}$ is the normalized $n$-th base station weight, which is computed at every iteration by
\begin{equation}
	\Bar{w}_{norm, n} = \frac{w_{norm, n}}{\sum_{n = 1}^{\textrm{N}} w_{norm, n}},
	\label{eq9}
\end{equation}
Equal weights are assigned to all the base stations during the first iteration. The uncertainty in each of the $\mathrm{N}$ position estimations computed by using each base station as a reference is evaluated by deriving the uncertainty factor, which is given by
\begin{equation}
	u_e = \frac{1}{\textrm{N} -1} \sum_{n \in bs\setminus \{e\}}^\textrm{N} |(\Delta{d})_{ne} - (||\hat{\textrm{Q}}_{\textrm{WA}}-\textrm{q}_{n}|| - ||\hat{\textrm{Q}}_{\textrm{WA}}-\textrm{q}_{e}||)|,
	\label{eq10}
\end{equation}
After deriving the uncertainties for all the position estimates, the weights for the next iteration are computed using the Andrews sine function, which is a robust function in the field of statistics and outlier rejection \cite{ANDREWS2015}.
\begin{equation}
w_{norm, n} = \begin{dcases*}
	  \frac{u_{max}}{u_{n} \Pi} \cdot \sin \left( \frac{u_{n} \Pi}{u_{max}} \right), & $\text{if } u_n \leq u_{max}$, \\
	0, & \text{else} \\
	\end{dcases*}  
	\label{eq11}
\end{equation}

where $u_{max}$ is the maximum acceptable uncertainty value beyond which the weights are assigned to $\mathrm{0}$, meaning that the measurement from the corresponding base station is rejected. At the end of each iteration, the convergence check is done between the current result and the previous result.   
\begin{equation}
	S = ||\hat{\textrm{Q}}^i_{\textrm{WA}} - \hat{\textrm{Q}}^{i-1}_{\textrm{WA}}||.
	\label{eq12}
\end{equation}
If the value of $S$ is found to be greater than a pre-defined threshold $\epsilon$, the algorithm starts repeating the process from the beginning. The workflow of IRLS is summarized in Algorithm~\ref{alg1}.

\begin{figure*}
	\centering
	
	\subfloat[\textrm{3D model}]{%
		\includegraphics[width=0.33\textwidth,valign=t]{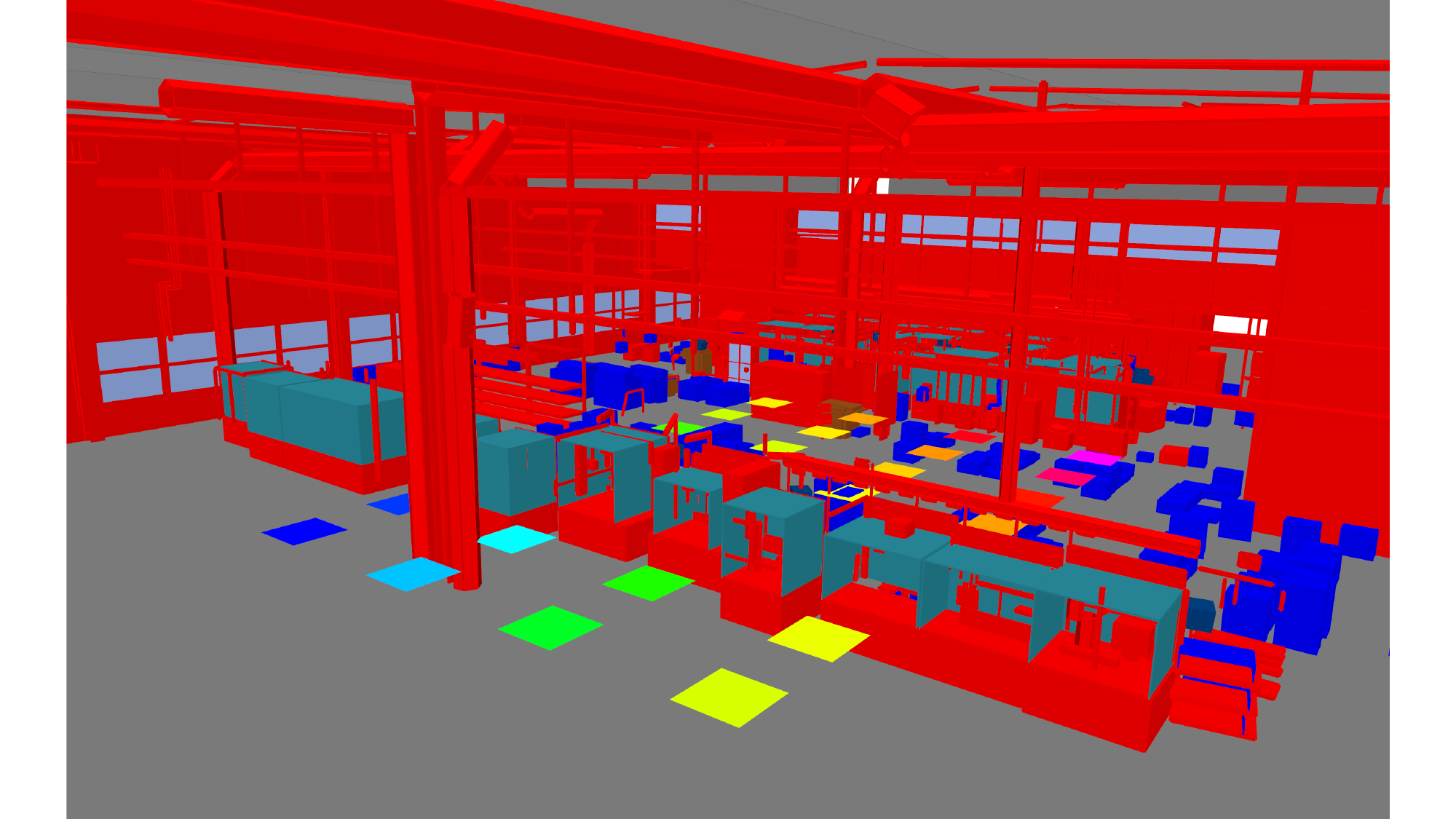}%
		\label{fig: 1a}
	} \quad
	\subfloat[\textrm{2D model}]{%
		\includegraphics[clip, angle=-90, trim=8.6cm 11cm 4.4cm 4.4cm, width=0.27\textwidth,valign=t]{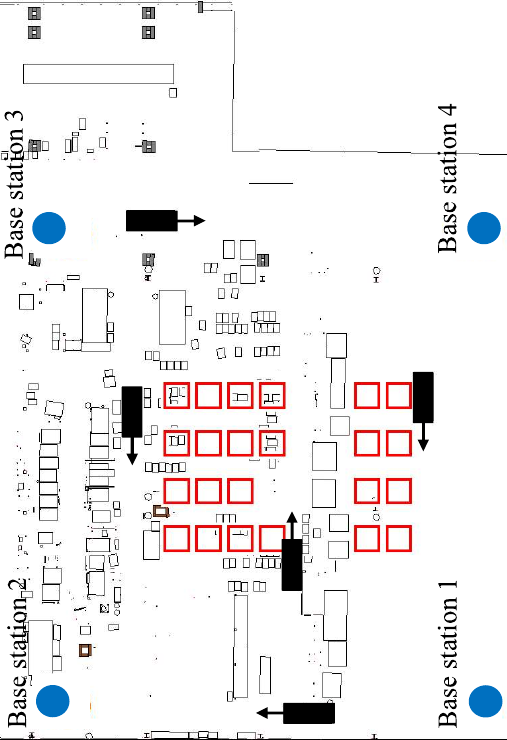}%
		\label{fig: 1b}
	} \quad
    \subfloat[\textrm{Multi-Path Components (MPCs)}]{
    	\includegraphics[width=0.33\textwidth,valign=t]{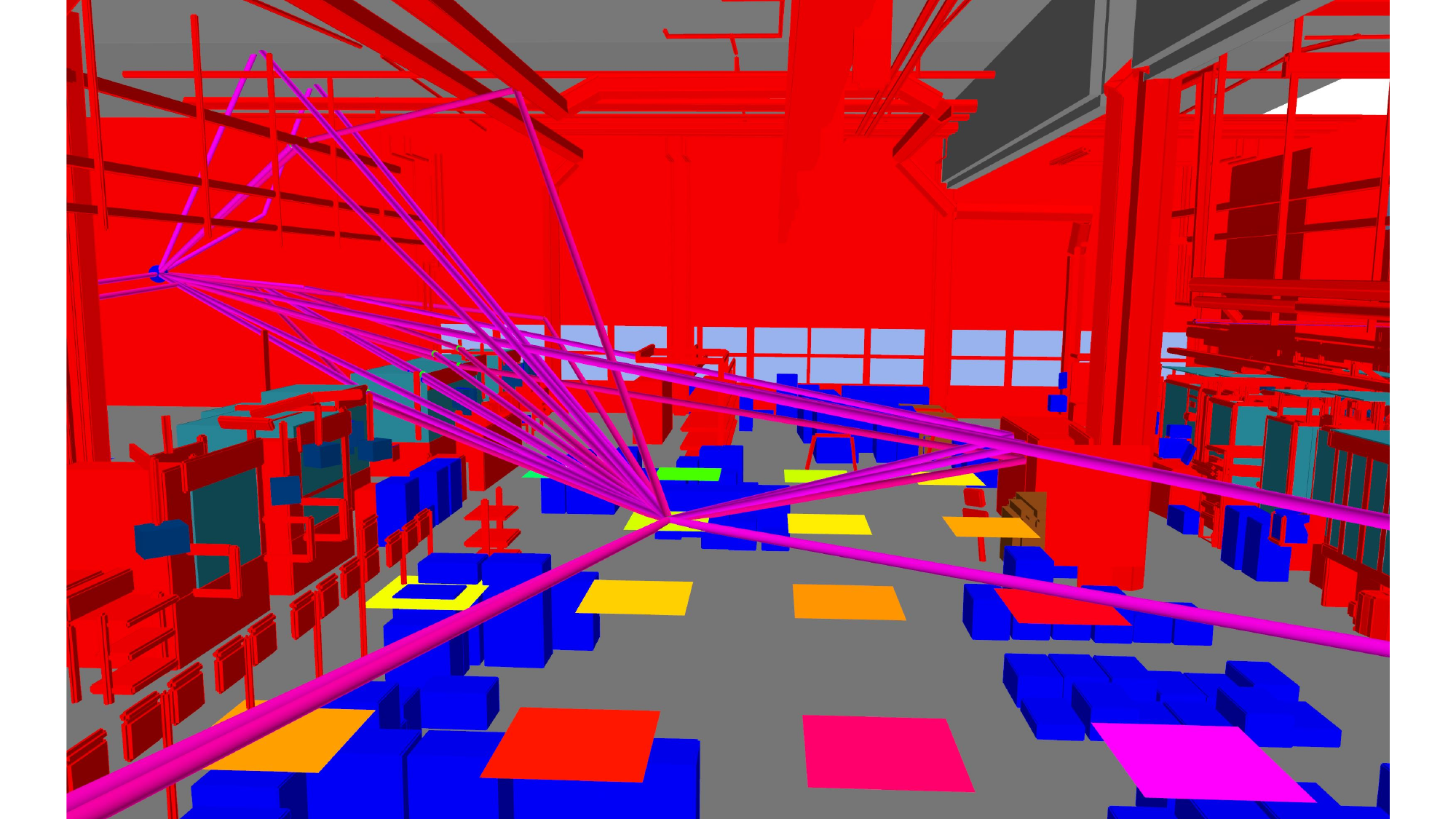}%
    	\label{fig: 1c}
    }
	
	\caption{Geometrical model of the industrial scenario. (a) The three-dimensional (3D) model of the production hall. (b) The two-dimensional (2D) model of the production hall with 4 base stations and 23 ground truth Points-of-Interest (PoI) (red squares). The black rectangles with the arrow mark indicate the forklifts and their moving directions. (c) The MPCs are emulated by the software between the base station and PoI pair.}
	\label{fig:1}
\end{figure*}

\section{Channel emulation in Ray-Tracing Enabled Industrial Scenario}
\label{Channel emulation}
We use the channel emulation technique Ray-Tracing due to its compatibility with Third Generation Partnership Project (3GPP) release-16 5G specifications for generating the ToA data for a dense-clutter real industrial scenario. The scenario corresponds to one of the production halls of Bosch with 42~m~$\times$~46~m~$\times$~8.8~m dimensions, located in Blaichach, Germany. The physical environment consisting of concrete walls, manufacturing lines, robot arms, and plastic storage boxes is scanned with 3-dimensional (3D) laser scanners, from which a Ray-Tracing enabled industrial model is created. Most of the objects in the industry are made up of metals. In addition, objects made with glass, wood, and plastic are also present. Material properties of these objects are defined based on the recommendation of the International Telecommunication Union (ITU) \cite{ITU2015}. The entire procedure of building an RT model can be found in \cite{HAN2022}. 

\begin{table}[t!]
	\caption{Altair Winprop settings for emulation of the channel.}
	\begin{center}
		{\renewcommand{\arraystretch}{1.4}
			\begin{tabular}[b]{ | p{3.4cm} | p{1.5cm}| p{1.5cm} |} \hline
				%\multicolumn{3}{|c|}{A Classification of Cats}\\ \hline
				
				\hfil Access method & \multicolumn{2}{|c|}{OFDM}\\ \hline
				\hfil Transmission scheme & \multicolumn{2}{|c|}{TDD}\\ \hline
				\hfil Use of frequency & \multicolumn{2}{|c|}{3.775 GHz / 26.85 GHz}\\
				\hline
				\hfil Use of bandwidth & \multicolumn{2}{|c|}{100 MHz / 400 MHz}\\ \hline
				\hfil Spacing of sub-carrier & \multicolumn{2}{|c|}{30 KHz / 120 KHz}\\ \hline
				\centering Antennas & \multicolumn{2}{|c|}{Omnidirectional}\\ \hline
				\hfil Transmit power & \multicolumn{2}{|c|}{20 dBm}\\ \hline
				\hfil Time for emulation & \multicolumn{2}{|c|}{60 s}\\ \hline
				\hfil Propagation & \hfil LoS & \hfil Yes\\ \cline{2-3}
				& \centering Penetration & 2nd order\\ \cline{2-3}
				& \centering Reflection & 2nd order\\ \cline{2-3}
				& \centering Diffraction & 1st order\\ \cline{2-3}
				& \centering Scattering & \hfil No\\ \hline \cline{2-3} 
			\end{tabular}
		}
	\end{center}
	\label{tab1}
\end{table}

The scenario of interest is shown in Fig.~\ref{fig:1}. The 3D model of the scenario can be seen in Fig.~\ref{fig: 1a}. The 5G positioning infrastructure consists of four base stations placed at a height of 4 m from the ground at four different corners of the Area of Interest (AoI), which spans roughly 29 m $\times$ 25 m. The choice of using this AoI for positioning comes from the fact that it is the initial priority area for enabling the 5G positioning service due to the presence of a supermarket, an area from where the Automated Guided Vehicles (AGVs) pick up the manufactured items. In this work, we define 23 static Points of Interest (PoI) of height 1 m, shown by red squares in Fig.~\ref{fig: 1b}. The algorithm performance is evaluated at these ground truth PoIs. For evaluation, two different emulation scenarios have been defined and studied. First, a static scenario involving just the static PoIs without any dynamic behavior in the scenario. Second, a semi-dynamic scenario, involving static PoIs and dynamic behavior in the form of moving forklifts represented by black rectangles in Fig.\ref{fig: 1b}, causes frequent NLoS.

A commercially available radio planning software Altair Winprop is used for the Ray-Tracing channel emulation. It takes the 3D model of the industry as well as the emulation settings listed in Table~\ref{tab1} as inputs to emulate the ToA data \cite{KARTHIKM2023}. For instance, Fig.~\ref{fig: 1c} shows the number of emulated signal paths (MPCs) by the software between the base station and the PoI pair. From the list of received MPCs, we use the first-arriving MPC as the preferred link choice for position estimation. Thereafter, the ToA data corresponding to the first-arriving MPC is utilized by the IRLS positioning algorithm for computing UE position as explained in Section \ref{IRLS}.   
%\footnote{https://web.altair.com/winprop-telecom}

\section{Results and Discussion}
\label{Discussion}
In this section, we present and describe the positioning accuracy obtained using the IRLS positioning technique in our previously described Ray-Tracing-enabled industrial scenario. The obtained results w.r.t. average positioning error and the 90th percentile are summarized in Table~\ref{tab2}.

\begin{figure*}[t!]
	\centering
	
	\subfloat[\textrm{LS positioning.}]{%
		\includegraphics[clip, trim=5cm 0cm 8cm 6cm, width=0.45\textwidth, valign = t]{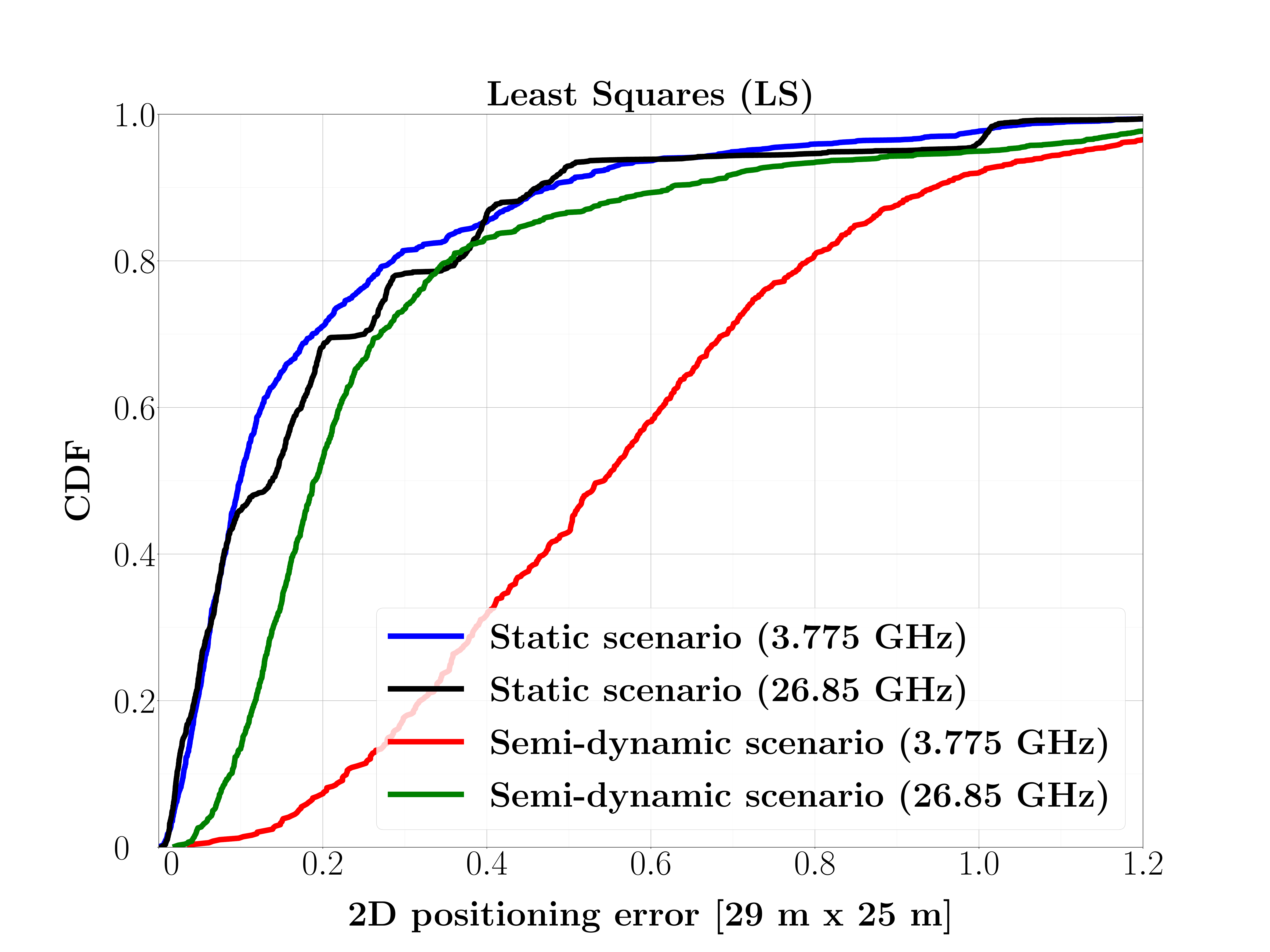}%left,bottom,right,top
		\label{fig: 2a}
	} \quad
	\subfloat[\textrm{IRLS positioning.}]{%
		\includegraphics[clip, trim=5cm 0cm 8cm 6cm, width=0.45\textwidth, valign = t]{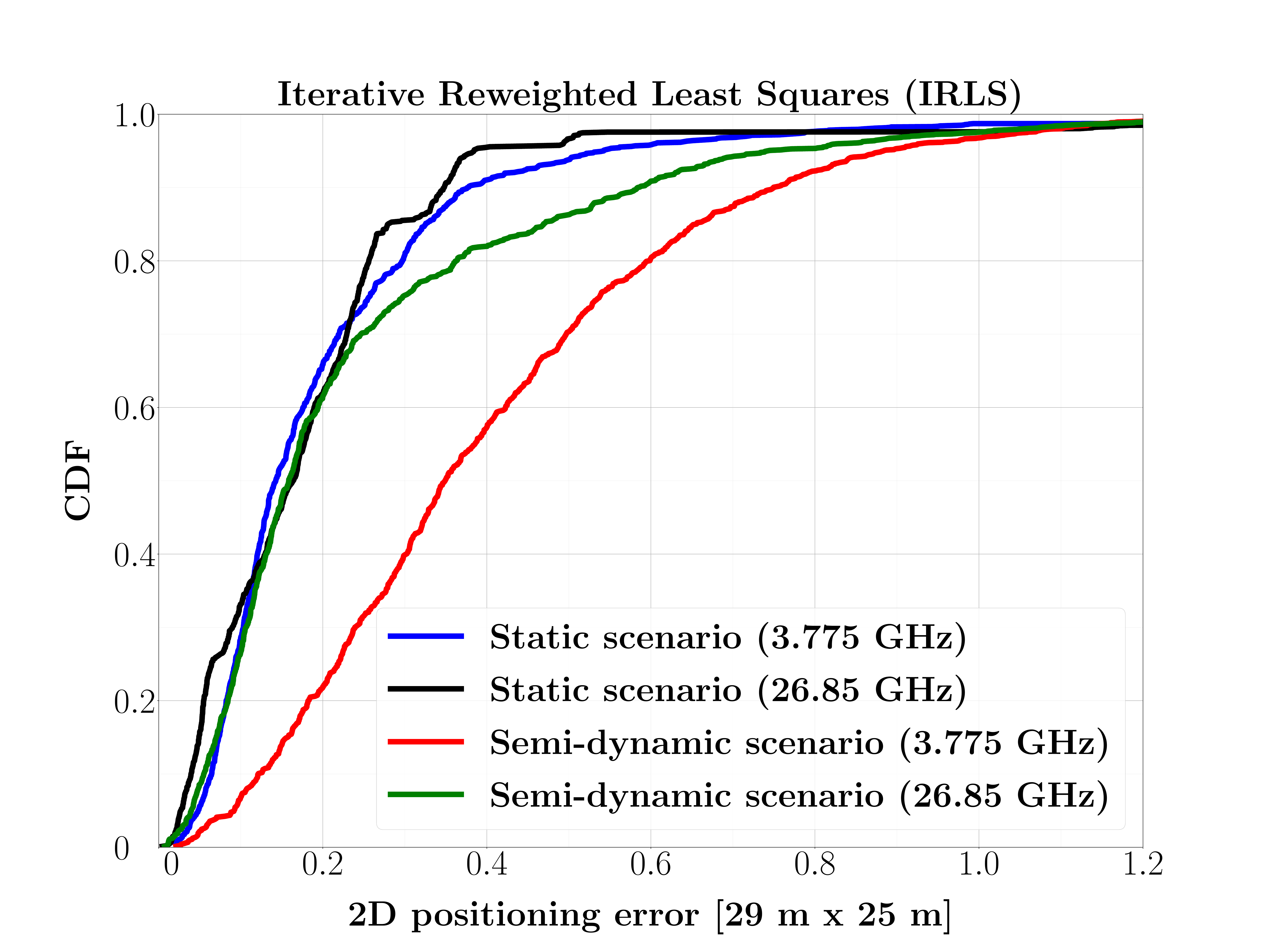}%
		\label{fig: 2b}
	}
	
	\caption{The Cumulative Distribution Function (CDF) of 2D positioning error for static and semi-dynamic scenarios under both the C-band and the mmWave-band. (a) Performance of LS positioning. (b) Performance of IRLS positioning.}
	\label{fig: 1}
\end{figure*}

In our industrial scenario, the PoIs that are in the LoS with the base stations yield high accuracy gain. This is because the first-arriving MPC to these PoIs corresponds to the LoS path or the diffracted path, whose path length is similar to that of the LoS path. While, the PoIs that lie behind heavy machinery and pillars, yield low accuracy gain. For these PoIs, the LoS paths to the base stations are blocked by the industrial infrastructure. As a result, the first-arriving MPC to these PoIs corresponds to the NLoS paths (reflections), that are longer in path lengths. The additional path length error, introduced by the NLoS MPC, causes the UE to misinterpret the real distance to the base station. Under such circumstances, if the selected reference base station is in NLoS with the PoI, this can lead to inaccurate position estimation, thereby, degrading the overall positioning accuracy.

\begin{table}[t!]
	\caption{Positioning error.}
	\begin{center}
		\newcommand\T{\rule{0pt}{2.6ex}}       % Top strut
		\newcommand\B{\rule[-3.8ex]{0pt}{0pt}} % Bottom strut
		{\renewcommand{\arraystretch}{1.4}
			\begin{tabular}[b]{|p {1.8cm} |M {1.8cm} |M {1.3cm} |p {0.7cm} |p {0.7cm} |}
				\hline
				& {\textbf{Scenario setup}} & \multirow{2}{*}{\hfil{\textbf{\vtop{\hbox{\strut Frequency}}}}\B} & \multicolumn{2}{|c|}{\textbf{TDoA}} \\ \cline{4-5}
				& & & \hfil \textbf{IRLS} & \hfil \textbf{LS}\\ \hline
                %& {\textbf{Scenario setup}} & \multirow{2}{*}{\hfil{\textbf{\vtop{\hbox{\strut Frequency}\hbox{\strut band}}}}\B} & \multicolumn{2}{|c|}{\textbf{TDoA}} \\ \cline{4-5}
				%& & & \hfil \textbf{IRLS} & \hfil \textbf{LS}\\ \hline
				
				\multirow{2}{*}{\hfil \textbf{\vtop{\hbox{\strut Mean}\hbox{\strut positioning}\hbox{\strut error [m]}}}} &  \multirow{2}{*}{Static} & \hfil C-band & \hfil 0.24 & \hfil 0.45 \\ \cline{3-5}
				& & \hfil mmWave & \hfil 0.21 & \hfil 0.36\\ \cline{2-5}
				& Semi-dynamic & \hfil C-band & \hfil 0.66 & \hfil 0.83 \\ \cline{3-5}
				& & \hfil mmWave & \hfil 0.44 & \hfil 0.52 \\ \cline{1-5}
				\hline
				
				\multirow{2}{*}{\hfil \textbf{\vtop{\hbox{\strut 90th percentile}\hbox{\strut [m]}}}} &  \multirow{2}{*}{Static} & \hfil C-band & \hfil 0.38 & \hfil 0.48 \\ \cline{3-5}
				& & \hfil mmWave & \hfil 0.35 & \hfil 0.46\\ \cline{2-5}
				& Semi-dynamic & \hfil C-band & \hfil 0.75 & \hfil 0.94 \\ \cline{3-5}
				& & \hfil mmWave & \hfil 0.58 & \hfil 0.62 \\ \cline{1-5}
				\hline
				
				%\multicolumn{2}{c}{\multirow{2}{*}{Multi-col-row}}&X\\
				%\multicolumn{2}{c}{}&X\\
				%\hline
			\end{tabular}
		}
	\end{center}
	\label{tab2}
\end{table}

The positioning results obtained by using the LS technique without employing the outlier rejection technique are depicted in Fig.~\ref{fig: 2a}, which shows the 2D positioning error Cumulative Distribution Function (CDF) for static and semi-dynamic scenarios for both the C-band and the mmWave-band. Looking at the curves of the static scenario, one can note that, at the 90th percentile, positioning in the mmWave-band achieves a positioning error of 0.46 m, which is 4.1$\%$ smaller compared to the 0.48 m achieved by positioning in the C-band. The possible explanation for this behavior lies in the fact that the ToA resolution depends directly on the signal bandwidth of the C-band (100 MHz) or the mmWave-band (400 MHz), which affects the accuracy of the ToA estimates. Similar behavior can also be seen for the semi-dynamic scenario, where positioning in the mmWave-band achieves 0.62 m, which is 34$\%$ smaller compared to the 0.94 m achieved by positioning in the C-band. In the current scenario, due to the movement of forklifts in the background, the NLoS condition occurs more frequently and the available LoS paths between the base stations and the PoIs reduce, forcing each PoI to use the first arriving NLoS MPC for position computation. And that elucidates the rise in positioning error for the semi-dynamic scenario compared to the static scenario.  

The performance of the IRLS positioning technique and the importance of rejecting the outliers can be assessed in Fig.~\ref{fig: 2b}. In comparison to the LS positioning, by using IRLS we could avoid the larger positioning errors at higher percentiles for both the emulation scenarios. For instance, in a static scenario under the C-band and the mmWave-band, we could lower the positioning error roughly by 20.8$\%$ and 23.9$\%$ respectively. Similarly, in the semi-dynamic scenario, we could dwindle the positioning error to roughly 20.2$\%$ and 6.4$\%$ under the C-band and the mmWave-band respectively. In brief, we demonstrated that in our particular industrial scenario positioning in mmWave-band outperforms the positioning in C-band. In conclusion, the idea of identifying and rejecting the outliers beforehand can lead to improved positioning accuracy.

%For the static scenario, one can note by observing the curves at the 90th percentile that, positioning in mmWave-band achieves 0.38 m error, which is 14.2$\%$ smaller compared to the 0.39 m achieved by positioning in C-band. The possible explanation for this behavior lies in the fact that the ToA resolution depends directly on the signal bandwidth, which affects the accuracy of the ToA estimates. Similar behaviour can also be seen in semi-dynamic scenarios, where, positioning in mmWave-band achieves 0.57 m, which is 38.6$\%$ smaller compared to the 0.78 m achieved by positioning in C-band. Due to the movement of forklifts in the background, the NLoS condition occurs more frequently and the available LoS paths between the base stations and the PoIs reduce, forcing each PoI to use the first arriving NLoS MPC to compute the position. And, that explains the increase in positioning error for semi-dynamic scenarios. Therefore, in our emulation scenarios positioning in mmWave-band achieves high accuracy due to the availability of larger bandwidth that is favorable for ToA estimation. 

\section{Conclusion}
\label{Conclusion}
The robust positioning scheme called IRLS for identifying and rejecting the outliers in the ToA measurements has been applied in the dense-clutter Ray-Tracing enabled industrial scenario. The results show the improvement in the positioning accuracy, with the applied IRLS under both static as well as semi-dynamic scenarios. Specifically, when compared with the state-of-the-art LS technique, we are able to avoid larger positioning errors at the higher percentiles. In addition, the choice of using either C-band or mmWave-band for indoor positioning in industrial scenarios has its own advantages and disadvantages. The C-band has less attenuation but gives more multipaths. This results in more outliers in the signal measurements. On the other hand, the mmWave-band has more attenuation in NLoS situations and results in fewer multipaths. However, a high density of base stations is required in the mmWave-band to tackle the problem of signal attenuation and guarantee optimal positioning accuracy. In our specific industrial scenario with four base stations, positioning in the mmWave-band provides more accurate results compared to the positioning in the C-band, achieving an average error of 0.21 m and 0.44 m in static as well as semi-dynamic scenarios respectively. 
	
	\section*{Acknowledgment}
	This work has received funding from the European Union's Horizon 2020 research and innovation programme under the Marie Sklodowska-Curie grant agreement ID~956670. Also, this work is part of the project ROUTE56 with grant PID2019-104945GB-I00 funded by MCIN/AEI/ 10.13039/501100011033 and the project 6-SENSES with grant PID2022-138648OB-I00 funded by MCIN/AEI/ 10.13039/501100011033 and by ERDF A way of making Europe.

	\bibliographystyle{IEEEtran}
	\bibliography{IEEEabrv,references}
	
\end{document}